# A scalable approach to undergraduate research in physics

Amanda L. Baxter[1*], Rafael F. Lang[1], Craig Zywicki[2], Stephanie M. Gardner[3], Abigail Kopec[1,4], Andreas Jung[1]

**Affiliations:**

[1]Department of Physics and Astronomy, Purdue University, West Lafayette, IN 47907

[2]Office of Undergraduate Research, Purdue University, West Lafayette, IN 47907

[3]Department of Biological Sciences, Purdue University, West Lafayette, IN 47907

[4]Department of Physics, University of California San Diego, La Jolla, CA 92093

*Correspondence to: adepoian@purdue.edu

## Abstract

Course-based undergraduate research experiences (CUREs) increase students' access to research. This lesson plan describes an interdisciplinary CURE developed to be able to involve over 60 students per semester in original research using data from large particle physics experiments and telescopes, although the methods described can easily be adopted by other areas of data science. Students are divided into research teams of four, which greatly leverages the instruction time needed for mentoring, while increasing research productivity by creating accountability amongst the students. This CURE provides a strong framework, which minimizes barriers that students may perceive. This helps increase the number of students that benefit from a research opportunity while providing guidance and certainty. Through this CURE, students can engage in original research with the potential for publication-quality results, develop communication skills in various modes, and gain confidence in their performance as a scientist.

## Table of Contents

# 1. Scientific Teaching Context

## 1.1 LEARNING GOALS

Students will:

1. engage in original scientific research in physics data analysis.
2. apply classroom knowledge to cutting-edge research problems.
3. experience auxiliary research aspects such as proposal writing and presentations.
4. develop effective interpersonal communication skills.
5. gain confidence in their performance as a scientist.

## 1.2 LEARNING OBJECTIVES

Students will be able to:

1. analyze large physics data sets using Python.
2. plan and execute research using data from leading physics experiments.
3. state research questions and formulate hypotheses.
4. demonstrate a range of scientific communication skills and effectively work in a team.
5. draw research conclusions and communicate results.

# 2. Introduction

As defined by the Council for Undergraduate Research [1], undergraduate research is a "mentored investigation or creative inquiry conducted by undergraduate students that seeks to make a scholarly or artistic contribution to knowledge". At Purdue University, undergraduate students gain research experience on campus, typically within one of two models: as an apprenticeship (paid or for credit) or within a course (for credit). Apprenticeships are the traditional model [2], but these experiences are limited to a small number of students due to their traditional structures consisting of one mentor and one mentee. Also, higher competition over coveted apprenticeships may contribute to unintentional exclusion of underrepresented populations. This makes access particularly difficult for students early in their academic career.

Kuh & Schneider [3] identified undergraduate research as a high-impact practice in higher education, but also identified students' access to these practices as a key concern [3]. To address access to research experiences, the Purdue University Office of Undergraduate Research (OUR) established a faculty development program (based on CUREnet [4]) designed around two objectives: (a) in the immediate term, to provide instructors with knowledge and skills to integrate research experiences into a course, and (b) to reduce barriers for undergraduate students to participate in research experiences [5], targeted to make such opportunities available to more students, especially a more diverse set of students. These courses are formally known as course-based undergraduate research experiences (CUREs) [2]. The CURE model—when compared to an apprenticeship—has been shown to be a cost-effective way to increase access to research experiences for more students [5, 6].

Besides increased access to research experiences, other benefits of CUREs to faculty and students include [5, 8]:

- the ability for one instructor to mentor many students (and, as is the case here, engage graduate students as mentors)
- opportunities for students to collaborate in team-based research experiences
- students engaging in the entire research process
- support for the instructor's own research program

Impacts of CUREs in physics have been published, see [9] for a recent example, but despite a thorough literature search using the pertinent databases and recent reviews, we were not able to find published teaching materials describing the design of a Physics CURE.

## 2.1 RESEARCH TOPIC

This CURE provides students the opportunity to conduct original research via data analysis from particle physics experiments and telescopes. The research questions in this CURE are open-ended and fluid. This contrasts with research that is for example conducted in more traditional wet labs, where students follow a rigid procedure running a piece of equipment which produces known effects (e.g., [10, 11]). In contrast, here, the physics data analysis procedure cannot be planned in advance. Following a brief introductory period, the research question and steps are determined from week to week, only after seeing the results from the previous week. While this course is based on physics research, as discussed below, we provide guidance for how to adapt this model for other research areas in physics and beyond.

## 2.2 INTENDED AUDIENCE

This course is composed of mostly first-year students pursuing a degree in physics, computer science, math and statistics, and engineering, though there have been students from other majors and class-standings enrolled. The course is taught as part of the Purdue University Data Mine initiative [12]. Students who enroll in this course can satisfy a partial requirement for a Data Science certificate at Purdue University.

## 2.3 REQUIRED LEARNING TIME

This course is designed to be taught over the course of one semester, or 16 weeks. Each week, the course requires three hours of in-person class time, plus an additional six hours of time committed to working on research and assignments outside of class. This course continues for a second semester, but the first semester (PHYS 323: Research in Big Data I) course is not a prerequisite for the second semester (PHYS 324: Research in Big Data II). The two courses can be taken independently or in sequence. This work describes the entirety of Research in Big Data I, but Research in Big Data II has a lot of the same components. The lectures and final project are different between the two semesters. We encourage students to take both semesters to continue to grow and develop their research skills.

## 2.4 PREREQUISITE STUDENT KNOWLEDGE

This course is designed for students with no prior coding or physics knowledge. Students are required to have completed high school mathematics and science courses. Students should have an excitement and drive for research and learning. Of course, some more senior students bring knowledge from their majors to the table, which is considered when assigning the research teams.

## 2.5 PREREQUISITE TEACHER KNOWLEDGE

As this course is a research-based experience, instructors need a strong research background in the field of research which is used in this CURE. A research question with multiple subprojects, and relevant scientific data are needed for this CURE. In our experience, research topics that are non-time-critical, exploratory studies complementary, but detached from, the main science goals intended by the experiments are best suited for this course. Additionally, as this course relies heavily on mentoring student teams in their research projects, instructors and Teaching Assistants (TAs) require experience in analyzing the data that is given to the students.

## 2.6 REQUIRED RESOURCES

This CURE is based on dividing students into teams in order to conduct their research. Because of this team-based learning, a TA to student ratio of 1:4 is required. Additionally, there needs to be one instructor for every 1-4 cohorts.

The research conducted in this CURE revolves around the analysis of vast data sets ("big data") from particle physics experiments and telescopes. To conduct research in big data, computing resources need to be made available. This CURE used a computing cluster with 7 login nodes and 28 batch nodes, shared among other courses with high performance computing needs [13].

## 3. Course Structure

We found that a clear course structure is instrumental to help undergraduate students be productive with their original research [14]. Therefore, the following key components were created for this scalable CURE: research teams to create accountability to students and provide guidance; a weekly tutorial for each research team for intense research mentoring; a seminar where students present their work to their peers; and a lecture to teach students relevant concepts. Homework and a clear syllabus complement this structure. In our experience, this added structure makes students more productive researchers compared to individual apprentice-style experiences.

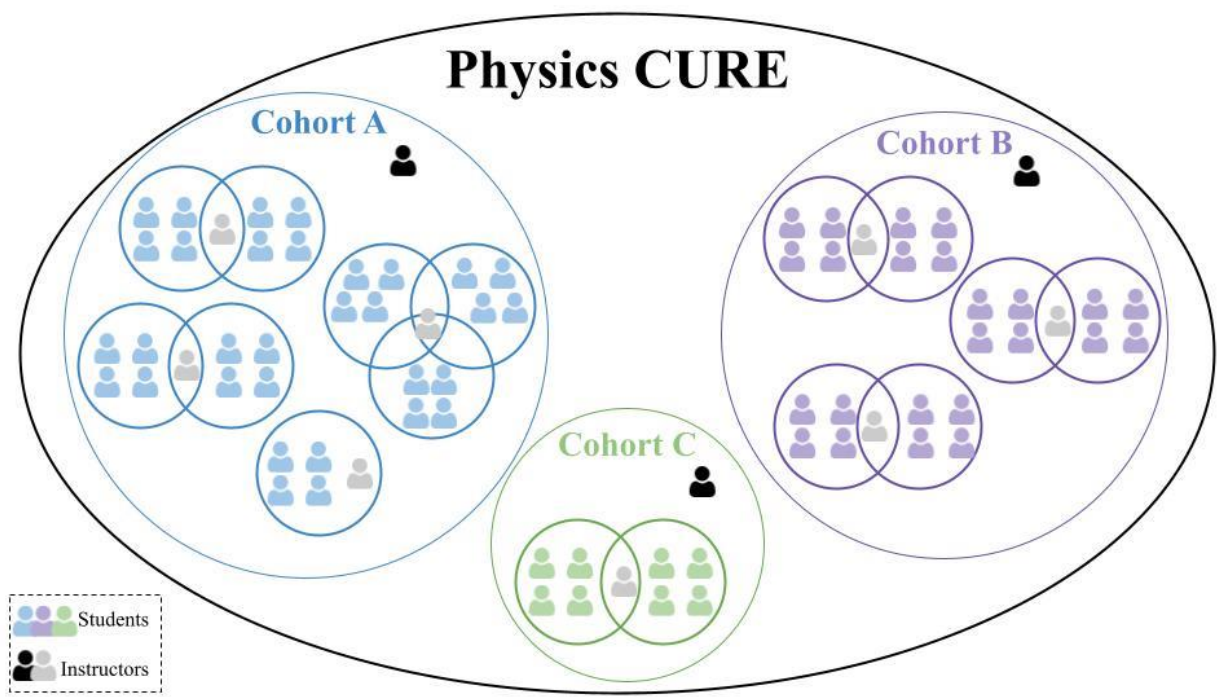

**Figure 1.** Course Structure. Diagram illustrating the structure of the CURE. The entire CURE shares the same lecture. This CURE was divided into three research cohorts, each led by a faculty instructor (black figure), working on different research areas, and having their own seminar. Each cohort was then broken down into research teams of four students (blue/purple/green figures), each supervised by a Teaching Assistant (grey figure) with their own weekly tutorial.

Students are assigned in teams of four to create accountability for students and reduce the mentoring load. This number is chosen based on experience, though we have found the literature to support this group size [15]. Odd numbers with groups of three and five were tried, but often suffered from an imbalance in the work done by the students. The instructor assigns the teams before the first day of class, with a special eye to pair members of minoritized groups into research teams to reduce perceived isolation. Further, the teams are assigned to diversify the students' backgrounds, mixing the different majors and seniorities to achieve possibly synergistic teams. Working in such teams encourages all students to actively participate and feel included in the CURE [16]. Additionally, we found that the students appreciate the opportunity to be assigned into teams, just like they would be in an eventual working environment. When students have trouble working with each other, we present this as an opportunity to try out team skills in an environment that has little impact on the students' life (unlike a similar situation in the workplace).

Tutorials are the main section where research teams work with their TAs to advance their research. Many teams working on the same overall research project are collected in one cohort, supervised by a faculty member that is an expert in this research area. Each cohort has a seminar where the students have an opportunity to communicate their progress to each other. Students thus can calibrate their performance and progress against other research teams, thus hopefully reducing imposter syndrome [17]. All cohorts come together for the lecture, where general content which is relevant to all physics research is taught. For the remaining time, students are required to schedule amongst themselves a time to work with their research teams. Figure 1 illustrates this course structure.

## 4. Scientific Teaching Themes

### 4.1 ACTIVE LEARNING

This course has multiple components which are taught differently. The lecture is designed to be more teacher-centered, with opportunities for group discussions. The seminar is designed to be student-centered. The students are giving presentations to their peers, and the instructor is never standing at the front of the room. The tutorial is designed to be bidirectional. The students are engaging with their TA and having an open dialogue and discussion in a group setting.

This course groups students in teams of four to conduct original research to engage in team-based learning [18]. Students are constantly communicating with their team, TA, and cohort about their research goals, progress, and challenges. Allowing for scientific inquiry gives the students the opportunity to try different approaches to solve their research question [19].

### 4.2 ASSESSMENT

Though students are broken into different research cohorts and teams, they all complete the assignments described below, regardless of their research topic. The course is heavily structured and scaffolded to provide the students with support and a sense of certainty throughout the semester.

**Homework** (Supporting Files S01-S12) – Students are assigned homework each week related to the topics covered in the lecture to reinforce these concepts. Homework ranges from conceptual questions, to calculations, to coding, and generating plots. Every homework assignment includes a reflection question where the students are asked to relate the lecture topic to their research question. This enforces the connection of the more general lecture content to their more specific research challenges. *(LG: 2&5)*

  **Physics Presentation** (Supporting File S13) – Each student presents an 8-minute presentation on a topic related to their research during the seminar. These are done in the style of a conference talk given at meetings of the American Physical Society. This assignment is designed to give students the experience of preparing a talk and communicating a relevant science topic to their peers. As this presentation is done during the seminar session, the cohort's instructor and TAs each grade the student's presentation. The grades are then averaged, and this becomes the student's final grade. *(LG: 1,3&4, LO: 3,4&5)*

  **Research Briefing** (Supporting File S14) – Each research team gives 2-3 status updates to their cohort during the seminar. These are evenly spaced throughout the semester to allow for feedback from the instructor and TAs to push the research forward. The students prepare slides, or a Jupyter Notebook [19], and update the cohort on their research question, progress, and next steps. These presentations are 10–15 minutes long, with ample opportunity for open dialogue and discussion. These research briefings are intended to replicate updates that a researcher would give in a research group meeting, which are less formal or polished than a typical conference setting. This assignment is also graded by all instructors and then averaged. Matrix grading (see below) is then applied to this averaged grade. *(LG: 1,2,3,4&5, LO: 1,2,3,4&5)*

  **Proposal** (Supporting Files S15-S19) – Securing funding is a relevant challenge for any academic career. Each research team is asked to write a proposal at the conclusion of the semester in the style of a grant proposal expected, e.g., by the National Science Foundation. After the first semester, students are in the position to propose a continuation of their research in the coming semester, independently of whether they enroll in the course again or not. The proposal is broken down into four parts (Team Preparedness, Literature Review, Preliminary Results, Proposed Work) which are assigned throughout the semester to force continuous progress. This also allows the students to receive and implement feedback before all the assignments are pieced together with proposed work for the final proposal. As all parts of this assignment are done as a team, and matrix grading is applied to each submission. *(LG: 1,2,3&4, LO: 1,2,3,4& 5)*

  The Team Preparedness assignment asks students to think about the skills they bring to the table and how the skills of each member of the team work together synergistically. The Literature Review asks students to reflect on the relevant literature related to their research topic. They should discuss the current progress towards the research goal and identify open research questions. Preliminary Results demonstrate that the research team has made progress towards their research question, is able to produce high-quality plots, and can discuss the implications of their research. Though the Proposed Work is due at the end of the semester, the assignment is designed for students to think about the unanswered questions and next steps of their work. The proposal should be written as if the students are asking for resources to continue their work for another semester.

  **Pre-Tutorial** (Supporting File S19-S20) – The first week of classes, students are required to schedule regular time to meet with their TA for the tutorial and schedule a few hours to meet with as a team to conduct their research. Before the team meets for the tutorial with their TA, they are required to inform the TA about any progress that was made during the last week. This can include any questions they have and issues they might be stuck with. This is an opportunity for the research team to reflect on what they did and what they are going to do. *(LG: 4&5, LO: 3&5)*

*Matrix Grading* (Supporting File S21) – Assignments conducted as a team effort (research briefings and all parts of the proposal) are graded using matrix grading, which we adapted from student grade weighing discussed in Ref. [16]. Matrix grading is applied to a percentage of the assignments total grade; applying it to the entire grade had a too-big effect. Students inform their TA what percentage they decide each member of their team contributed to the assignment beyond the student's own work. This penalizes students who don't participate or put in enough effort (in the eyes of their peers), and rewards students who carry the team. The goal is that for a group of four, each member will award the rest of their team each 33% so that everyone receives 100% of the graded assignment. If matrix grading comes back severely lopsided, the TA will initiate a conversation with their research team during tutorial to discuss team functionality. This open communication puts everyone's expectations out in the open and the team can realign themselves to be a more balanced and functional team. In our experience, less than 5% of research teams are dysfunctional but following an intervention by the TA (and sometimes the faculty), students appreciate the opportunity to experience and deal with difficult colleagues (as they might in the workplace) but in a safe environment with no risk of existential damage.

## 4.3 INCLUSIVE TEACHING

Apprenticeship-style experiences (such as summer research) contain many barriers, including lack of awareness and financial or personal limitations [20]. In contrast, this CURE is offered to any student at Purdue University with no prerequisites. This enables a diverse set of students with regard to major, gender, race, ethnicity, and class-ranking. Carefully assigning students to their research teams of four both leverages this diversity to the benefit of the research and encourages all students to actively participate [21].

Students within a cohort work on research projects that are synergistic, rather than distinct. Each cohort has one overarching question which we hope to answer during the CURE, and each team explores an individual part of that research question. This allows for more student-to-student interactions and brings teams together to work synergistically. In our experience, this helps to reduce imposter syndrome that students might encounter, as they can calibrate their performance against that of their peers. Thus, it allows students to recognize that slow progress is inherent in original research, as opposed to an issue with their individual performance.

# 5. Lesson Plan

## 5.1 STRUCTURE

The class is broken down into four parts, which each have their own timelines. Each week there is a 50-minute lecture for the entire CURE, a 50-minute seminar for each research cohort, a one-hour tutorial for each research team, and approximately six-hours spent on research and homework. Table 1 outlines the timelines for the lecture and seminar and when various assignments are due. Figure 1 shows the CURE structure with the approximate number of students in a recent iteration of this CURE at Purdue University.

**Lecture** – Lecture for all students is scheduled for 50 minutes. The professor teaching the lecture tries to engage students using various aspects of the affective domain, as is described in Ref. [21]. Topics covered in lecture range from basic coding and presentation skills to statistical methods and particle physics. Following the lecture, students are to complete a short homework assignment which reinforces the concepts. It also asks students to reflect on the lecture and describe how the lecture topic relates to their research projects, in order to link the more general lecture topics to the more specific research problems.

**Seminar** – Seminar is scheduled for 50 minutes, following lecture, in which students are split into their research cohorts (8-40 students/cohort). This course was taught with three research cohorts, though the structure presented here is entirely agnostic to the number of cohorts: more or less cohorts can be used depending on the availability of research fields. Each cohort includes one faculty and multiple TAs.

The first weeks of seminar are led by the faculty to introduce specifics of the research field, the cohort's research question, and the initial research questions for the individual teams. After this, the seminar is a reverse-style environment, where students lead discussions about their research or present pertinent science topics. The seminar is designed to allow students to communicate their work to the other members of their cohort. This is done with two different assignments: one being the Physics Presentation (S13) and the other being the Research Briefings (S14).

**Tutorial** – The tutorial is scheduled for one hour during the week, where a research team (four students) meets with their TA. Each TA can oversee one or more research teams; typically, they mentor two. Tutorials are where there the research is pushed forward. Before the tutorial session, research teams are required to share their pre-tutorial assignment to their TA. During tutorials, research teams will update their TA with progress made during the previous week. The TA will answer any questions, from conceptual understandings to data interpretations to coding errors/bugs and guide the students further along their research.

## 5.2 TIMELINE

**Weeks 1-4** – During the first weeks, students are taught the basics to presentations, coding, and their research topic. With no background knowledge required of students, these first few weeks are crucial to their success in the course.

For an introduction to effective presentations, a lecture is given to the students that is broken down into four parts: Know your Audience, Be Understood, Be Prepared, and Slide Design. Students are most familiar with slides prepared for teaching, but in academia there are many types of presentations (i.e., colloquium, seminar, conference) which require different preparation. So, students need to be taught how to prepare talks for these various presentation types.

To get an overview of the research topic, two seminars are used for teaching the students (a) the working principle of the experiments and (b) open research questions and possible research topics. These lessons are also reinforced during tutorials, where TAs explain them again as needed.

For coding introductions, students are given two Jupyter Notebooks [18] as tutorials to getting familiar with Python and big data analysis. The first is a modification of "A Crash Course in Python for Scientists" developed by Rick Muller [23]. The second is a tutorial on data analysis techniques developed by the Purdue research team. The tutorial includes information on how to make various types of plots and data selections. After the students have practiced with these tutorials, they are given a lecture on Python Tips and Tricks. This lecture is mostly to teach students how to write code that is effective and readable, and how to visualize data. We found it beneficial to provide students with a working example analysis code to get them going on original research rather than basic coding issues.

**Weeks 5-10** – The lectures in the middle of the semester consist of topics related to data analysis, fitting, and statistics. By this point in the semester, students are working on their assigned research topic and understand how these data science topics are relevant to their research. In the seminar session, physics presentations have begun.

This part of the semester is past introductory material, and students are getting into the bulk of their research. Hence, in weeks 9 and 10, teams have their first research briefing. The first research briefing provides students with an opportunity to communicate their work to the rest of the cohort, see how the research projects are related to one another, and receive feedback and directions forward in their work.

**Weeks 11-16** – In the last third of the semester, the lectures transition into more topical discussions as instructors from each cohort give an overview of their research to the entire class. These lectures contextualize the students' research within the current state of the respective field and open questions that are driving the research community.

In the seminar, physics presentations and research briefings continue. The research progress at this point in the semester is more about wrapping up the current work than it is about trying new things. This period is where we see students be the most productive. The conclusions they make in the last third of the semester tend to be the more useful results obtained towards the cohort's research question.

The class is designed to lean more on instruction in the beginning of the semester and shift the focus more and more towards research over time, see Figure 2. At the start, students have little background in particle physics or astrophysics, big data science, or research, so they need the most instruction to build their knowledge base. As the semester progresses, students require less instruction and more time to work on research. By the end of the semester, most research teams are self-sufficient. Students contribute original results to the open questions posed to them at the beginning of the semester.

**Table 1.** Semester Timeline. Outline of lecture topics, seminar topics, and assignment due each week throughout the semester. Bold items indicate assignments which are completed as a research team and are partially graded using Matrix Grading.

| Week | Lecture Topic | Seminar | Assignments Due |
|---|---|---|---|
| 1 | Introduction, Syllabus, Assignments | Research Introduction I | Homework 1: Getting Started, **Homework: Team Scheduling** |
| 2 | Scientific Talks and Data Presentation | Research Introduction II | Homework 2: Good Presentations |
| 3 | No Lecture – Holiday | No Seminar – Holiday | **Team Preparedness** |
| 4 | Python Tips and Tricks | Physics Presentations I | Homework 3: Python 101 |
| 5 | Introduction to LaTeX | Physics Presentations II | Homework 4: Latex |
| 6 | Poisson Statistics and Gaussian Distributions | Physics Presentations III | Homework 5: Poisson Distributions |
| 7 | Fitting, $\chi^2$ | Physics Presentations IV | Homework 6: Fitting, **Literature Review** |
| 8 | No Lecture – Holiday | No Seminar – Holiday | *Optional Homework: Feedback (S22)* |
| 9 | Estimations and the Fermi Problem | **Research Briefing I** | Homework 7: Fermi Problem |
| 10 | Error Propagation, Estimating Systematic Errors | **Research Briefing I** | Homework 8: Statistical and Systematic Errors |
| 11 | Topical Lecture – Time-Domain Astrophysics | Physics Presentations V | Homework 9: Telescopes, **Preliminary Results** |
| 12 | Topical Lecture – Artificial Intelligence in Big Data | Physics Presentations VI | Homework 10: Artificial Intelligence |
| 13 | Topical Lecture – Hunting for New Physics with the LHC | Physics Presentations VII | Homework 11: Colliders |
| 14 | No Lecture – Holiday | No Seminar – Holiday | None |
| 15 | Topical Lecture – The Quest for Dark Matter | **Research Briefing II** | Homework 12: Dark Matter |
| 16 | Why Physics Beyond the Standard Mode | **Research Briefing II** | **Full Research Proposal** |

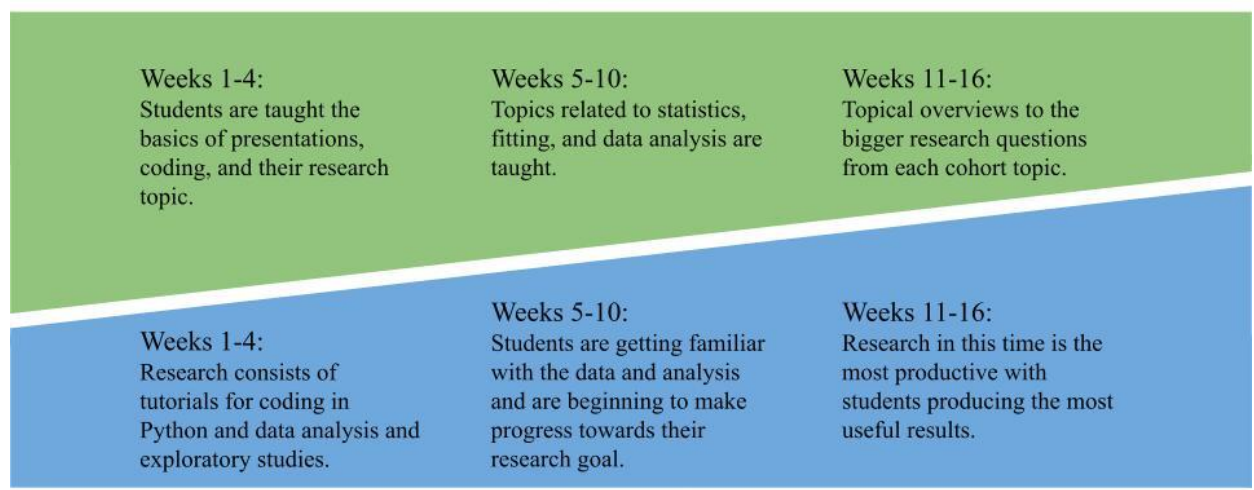

**Figure 2.** Instruction and Research Ratios. Diagram illustrating the amount of instruction and research that happens over the course of the semester. Instruction is front-loaded and decreases over the semester. While the research is minimal at the beginning of the semester and increases over the course of the semester, where the most progress is made in the last few weeks.

## 6. Teaching Discussion

### 6.1 GRADES

The grading scale for this course is skewed towards the high end (Table 2). It is relatively easy to get an A when the student puts in effort. We do not grade on the research achievement, as this is not under the control of the student, but for its presentation and communication. Being able to effectively communicate research progress and results and work in a team is one of the main learning objectives (LO: 5). The assessments are designed to reward students for success in these areas. Additionally, in this course, the importance of homework is to learn the material. If a student receives a poor grade on an assignment, they can take the feedback they received, redo the assignment, and resubmit it for a new grade.

**Table 2.** Grading Scheme. All assignments in the course add up to 1000 points. Based on the student's total points, the grade distribution is the following.

| >900 | 800-989 | 700-799 | 650-699 | 600-649 | 550-599 | 500-549 | 450-499 | 400-449 | 350-399 | 300-349 | <300 |
|---|---|---|---|---|---|---|---|---|---|---|---|
| A+ | A | A- | B+ | B | B- | C+ | C | C- | D+ | D | F |

## 6.2 LESSON EFFECTIVENESS

We can document the effectiveness of this CURE through two assessments. The first is the research briefings: Seeing how students' communication and progress changes from the first to the second research briefing gives us insight to the effectiveness of the lessons. Typically, we find that during the first research briefing, students jump right into the details of their work. What they often lack is communicating the big picture and their research question. But during the second research briefing, they do communicate the bigger question and bring their peers up to speed, before going into the details of their research. We find that later in the semester, more discussions arise between the research team and the instructors, and among the research teams. This shows that students not only engage in original research, but improve their science communication skills, and gain confidence to discuss their work with others.

Additionally, the final proposal builds throughout the course. A part of the assignment is due approximately every four weeks, see Table 1. The students receive feedback after each part and can make changes, before piecing all parts together for the final proposal. Thus, the final proposal documents their growth over the course of the semester and demonstrates their ability to conduct and communicate research.

A third way to determine the lesson effectiveness is by observing the progress of the research. Some projects continue as independent studies for undergraduate students, or a graduate student or postdoc might continue investigating the research question. Through this class, significant science results have been achieved. From the initial work of undergraduate students in this work, research progress was continued by a graduate student and postdoc and has resulted in multiple papers published [24–27] or in preparation. Additionally, several students from this course continue to do research in the group as independent studies or more apprenticeship style experiences.

## 6.3 SUGGESTIONS FOR IMPROVEMENTS OR ADAPTATION

There are many ways to adopt this course for various uses and research projects. We have not used the same research question two years in a row. The projects typically continue from Fall to Spring, but after this, enough progress has been made on the topic that it can become hard to find tasks for multiple undergraduate teams to work on. As the various experiments continue to take data, there is always new data to analyze and new theories to test out, thus the projects are ever evolving.

During the second semester, the course is modified from what is described here. The lectures cover topics from particle physics and data science. Additionally, the final project is changed. Instead of writing a proposal, students either present a poster at an undergraduate research symposium or they write a paper in the style of Physics Review.

At Purdue University, this course has not been formally approved to satisfy some of the university's and college's Core Requirements; this approval is pending will further enhance the utility of this course to our students.

## 6.4 PLANNING

Physics presentations are 8 minutes, plus 3 minutes for questions, which allows for 4-5 presentations in an hour-long seminar. This means 6-7 seminars are needed for physics presentations for a cohort of 30 students. Taking out holidays, this leaves four weeks for Research Briefings. With fewer physics presentations (i.e., fewer students in the cohort), more research briefings can be built into the course. A minimum of two research briefings are necessary. The first research briefing should be a minimum of four-weeks into the semester. During the first four weeks, students are learning how to code, make basic plots, and how the detector works. Up until this point, they are making only limited progress on their assigned research question. But by Week 9, students will have made a lot of progress, which allows for great discussions during the Research Briefings.

## 6.5 SCALABILITY

This CURE was conducted with research cohorts of various sizes (between 8 and 40). A main factor limiting scalability is the ratio of instructors/mentors to students. Since this CURE was conducted at a large research institution, TAs improved the ratio and allowed for large overall course enrollments (with over 60 students). Using research teams improves engagement with and mentoring of students; we recommend using teams up to the count manageable by the number of TAs.

# 7. Research Results

It is unlikely that the research project assigned to students will be concluded by the end of the semester. This expectation/realization is discussed at the beginning of the semester. Though the course is 16 weeks, students spend six hours a week on this course outside in-person meetings. But these six hours include homework, team assignments, and research. So realistically, each student spends maybe 3-4 hours on research each week. Over the course of the whole semester, which is less than two weeks full time per student. We thus make sure to clearly define the expectations of research from the beginning.

This CURE had up to four research cohorts and has been taught for eight semesters. During this time, there has been a lot of progress made in the various fields. Students in the CURE conducted exploratory studies, which were then followed up by graduate students and postdocs in the respective cohorts. One Cohort used data from the XENON direct detection dark matter experiments [28, 29]. Based on such exploratory research, more detailed studies resulted in two papers which are published [24, 27] and two which are in preparation [30, 31]. Another Cohort used data from the Compact Muon Solenoid (CMS) experiment, which is a high energy particle physics experiment at the Large Hadron Collider (LHC) [32]. The research done in this Cohort has been used in the work described in two upcoming publications. Further, one Cohort worked with data from the Zwicky Transient Facility, which is an optical time-domain survey [33]. This work resulted in one publication [25], and two additional papers, one under review [26] and one in preparation, which use results from the original research conducted by the undergraduate students.

## 8. Conclusions

We have designed and implemented a scalable approach to undergraduate research, enabling an original research experience in physics data analysis to more than 60 students per semester. This course can be easily adopted to other areas of physics that involve big data sets. Key points of this CURE are:

- research in teams of four, which reduces the mentoring load by that factor four, while creating a sense of accountability among the students, aiding in teaching students skills relevant to teamwork
- a strong structure and scaffolding, with formalized meeting times and assignments that help guide the students along and help students not getting overwhelmed by the open-ended and complex research tasks.
- a seminar where the students take charge of communicating their science with their peers, to help improve their communication skills and fight imposter syndrome.
- a truly original research experience, where the outcome from research steps in one week can change the tools needed in the following week, yielding results which have the potential to lead to peer-reviewed publications.
- the use of data sciences which through its low cost, low risk, and easy access make this CURE easily scalable to dozens of students.

This paper illustrates a course-based undergraduate research experience taught at a large research institution. This CURE (a) includes a novel research experience; (b) engages students' creativity and ownership in the discovery process and presentation; (c) is mentored by a Purdue faculty/staff instructor, and (d) is offered for credit. While many courses contain some of these characteristics, the inclusion of all four characteristics distinguishes this CURE from other experiences.

Over the last few years, this CURE has included more than 500 students from diverse backgrounds and across majors and provided them with the opportunity to conduct original research. The research questions stemmed from particle physics experiments and telescopes, which collect petabytes of data each year. In the era of big data science, the students were able to gain invaluable skills as they continue in their academic careers and beyond.

## Supporting Materials

- S01. Homework 1
- S02. Homework 2
- S03. Homework 3
- S04. Homework 4
- S05. Homework 5
- S06. Homework 6
- S07. Homework 7
- S08. Homework 8

- S09. Homework 9
- S10. Homework 10
- S11. Homework 11
- S12. Homework 12
- S13. Physics Presentation Rubric
- S14. Research Briefing Rubric
- S15. Team Preparedness Rubric
- S16. Literature Review Rubric
- S17. Preliminary Results Rubric
- S18. Proposed Work Rubric
- S19. Pre-Tutorial Assignment
- S20. Homework – Team Scheduling
- S21. Matrix Grading Example
- S22. Homework - Feedback

# Acknowledgments

Development of this course was in part sponsored through a Purdue University CURE Fellowship and the Purdue University Department of Physics and Astronomy. We thank the Purdue University Office of Undergraduate Research for the support in developing this course. We thank Alec Habig for useful comments on this manuscript.

# S01: Homework 1 - Getting Started

## 1. Jupyter Notebooks

We need to make sure everybody can run Jupyter by themselves right away, which is why we have this here as a separate assignment: Create a simple "hello world" beginner's program using a Jupyter notebook, greeting your TA with a personalized message.

**(10 points)**

## 2. Histograms

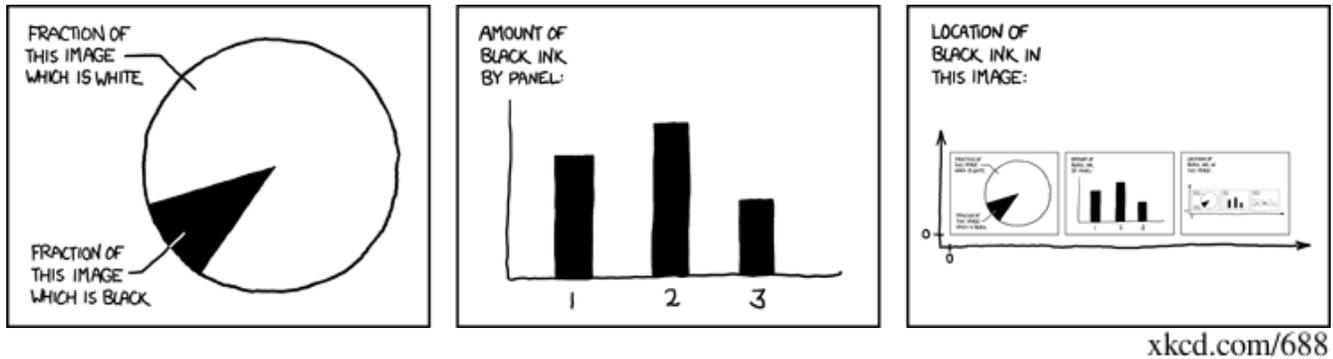

Histograms are a key visual to understand data. Use Jupyter together which whatever data you like.

a) Search around for online resources on python. Use your favorite resource to learn how to code a histogram in python, and thus submit a 1-dimensional histogram including proper axis labels.

**(10 points)**

b) Under which circumstances does it make sense to use logarithmic scaling on an axis?

**(5 points)**

## 3. Reflection

Every homework will have a reflection part, where we tie the content from the lecture to our research. In the first lecture, we talked about this being a course which features original research. How is this different from e.g., a typical lab course, even when that lab is well-designed and inquiry-based?

**(5 points)**

# S02: Homework 2 - Python 101

## 1. Functions

Modular code greatly improves readability and robustness. The key is to define clear boundaries between code parts with well-defined inputs, outputs, and a carefully chosen name for that function. This code draws a cylinder in 3D space, just copy and paste it into a Jupyter notebook:

```python
import numpy as np
import matplotlib
import matplotlib.pyplot as plt
import matplotlib.colors as clr
from mpl_toolkits.mplot3d import Axes3D
#Plot inline
%matplotlib inline

#initialize a 3D figure
fig = plt.figure(figsize=[6,5])
ax = fig.add_subplot(111,projection='3d')

#Define parameters of the cylinder (position, size)
x_cm = 0
y_cm = 0
z_cm = 0
radius_cm = 2
height_cm = 4

#Calculate cylinder surface, using code picked off of stackexchange
Zlist_cm = np.linspace(z_cm-height_cm/2,z_cm+height_cm/2,20)
thetalist_rad = np.linspace(0,2*np.pi,20)
thetagrid_rad,Zgrid_cm = np.meshgrid(thetalist_rad,Zlist_cm)
Xgrid_cm = radius_cm*np.cos(thetagrid_rad)+x_cm
Ygrid_cm = radius_cm*np.sin(thetagrid_rad)+y_cm

#Plot the cylinder using matplotlib
ax.plot_surface(Xgrid_cm,Ygrid_cm,Zgrid_cm,alpha=0.5,color='blue')
ax.set_xlabel('X (cm)')
ax.set_ylabel('Y (cm)')
ax.set_zlabel('Z (cm)')
#Set viewing angle
ax.view_init(15,-95)

plt.show()
```

Create a function that can be used to draw many cylinders quickly and easily without having to copy and paste the whole code each time. Submit a notebook with your function and an example that uses it to draw a bunch of cylinders in some arrangement you define.

**(15 points)**
**(continued next page)**

## 2. Aliasing in Histograms

If your data comes in discrete values, then histograms are prone to aliasing. Here is a simple example to illustrate the issue. How can you recognize that a histogram shows an aliasing artifact? What do you need to change in the plt.hist line to prevent aliasing? How can you implement this in general for any integer-valued data?

**(10 points)**

```
In [1]: import numpy as np
        import matplotlib.pyplot as plt
        np.random.seed(42) # I like my random numbers reproducible...

        # generate some data to play with
        mean = 61.5
        sigma = 16.9
        number_of_entries = 1234567
        data = np.random.normal(mean, sigma, number_of_entries)
        plt.hist(data, bins=100, range=(0,120), color="blue")
        plt.show()
```

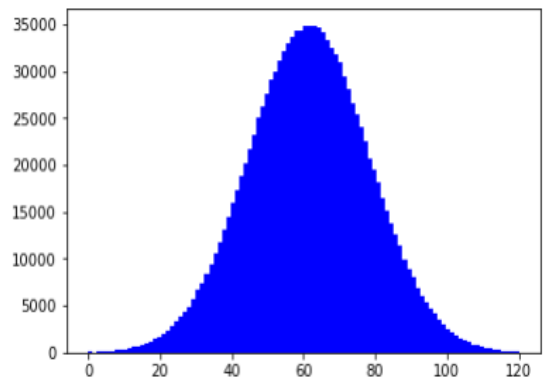

```
In [2]: # but what if data comes only with discrete values?
        data = np.round(data,0)
        plt.hist(data, bins=100, range=(0,120), color="red")
        plt.show()
```

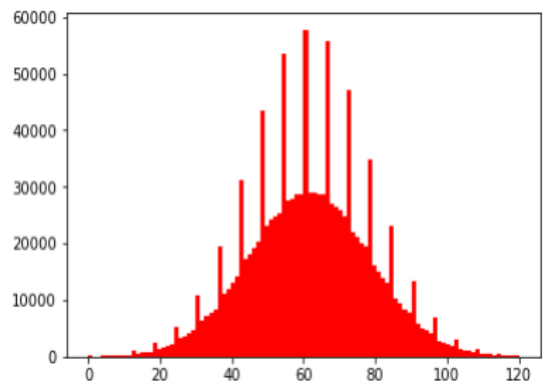

## 3. Reflection

How will this week's lecture content help you with your research project?

**(5 points)**

**S03: Homework 3 - Good Presentations**

**1. Scientific Presentations**

The internet is full of tips and tricks for good presentations, and more specifically, good scientific presentations. Search for trusted resources and see what you can learn from them. Submit the three most important lessons for you personally. Also cite the respective source that you used for each of them and note why you found it to be trustworthy.

**(25 points)**

**2. Reflection**

What is your personal take home message from this lecture?

**(5 points)**

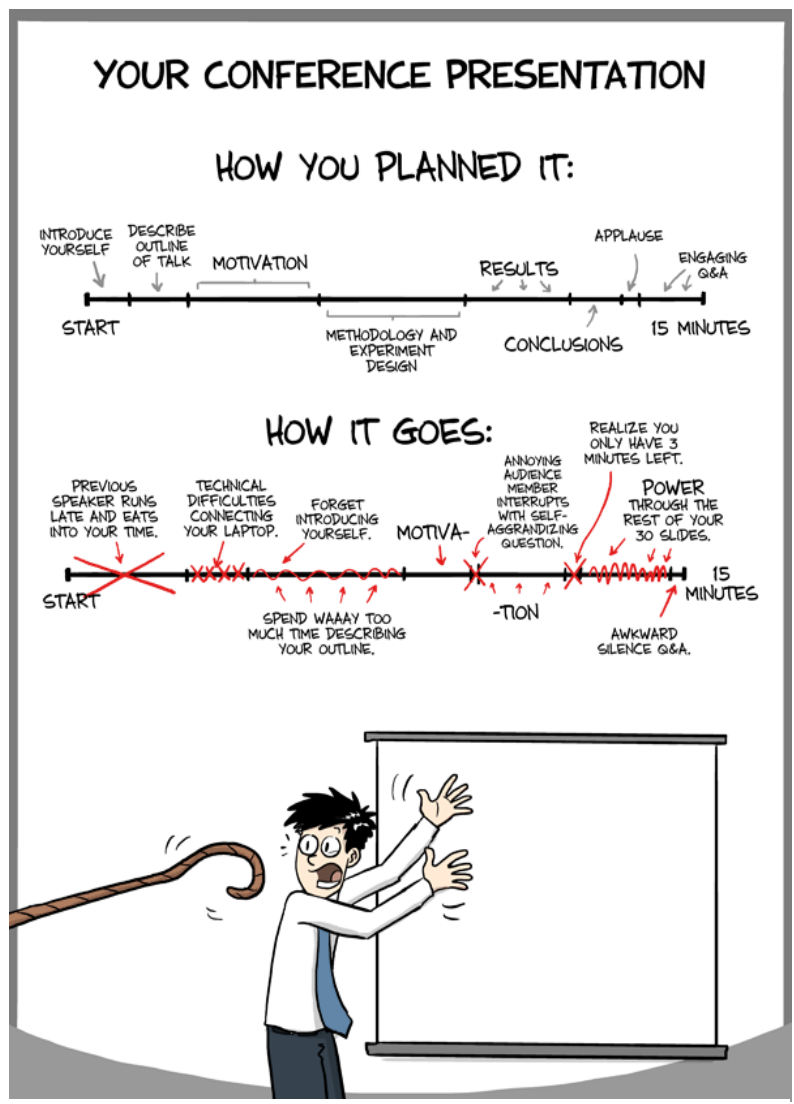

**S04: Homework 4 - Latex**

1. **Software**

This week we talked about "what you see is what you get" editors and latex. Which software do you currently use for each of these tasks? Do you intend to try out something else in the near future?
- Formal letters
- Applications packages including CVs
- Lab reports
- Presentation slides

**(25 points, graded completion only)**

**2. Reflection**

Where can fitting be important in your team's research project?

**(5 points)**

# S05: Homework 5 - The Poisson Distribution

No measurement is complete without reporting the associated uncertainty, which can be both statistical and systematic in nature. Two distributions are by far the most important to describe statistical errors: The Poisson distribution for measurements that yield discrete values (i.e., result in an integer number), and the Gauss distribution for measurements that yield continuous values. Here, we'll play with the former.

## 1. Poisson Process

Imagine some measurement of a discrete parameter $n$ subject to statistical fluctuations. For the sake of a silly example, let's consider the number $n$ of chipmunks you encounter during any given hour on campus. The distribution of chipmunks you will encounter is given by a Poisson distribution,

$$Poisson(n) = \frac{\lambda^n e^{-\lambda}}{n!}$$

where $\lambda$ is the *expectation value* or *average* of the distribution.

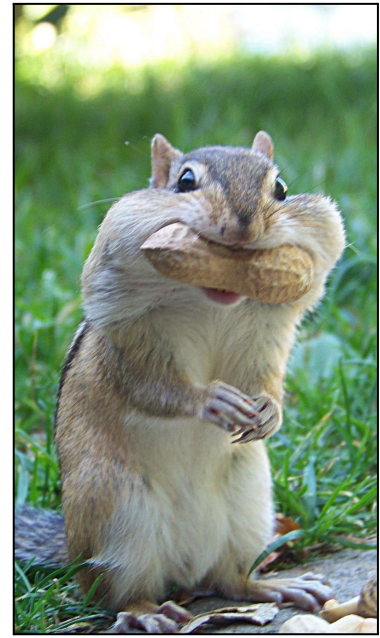

a) Write a Jupyter note to calculate the probability of observing $n$ chipmunks given an expectation value $\lambda$.

**(10 points)**

b) Suppose you spend your one-hour lunch breaks this week doing something useful, by sitting near the engineering fountain counting chipmunks. You observe the following:
        Monday: 3 chipmunks
        Tuesday: 6 chipmunks
        Wednesday: 1 chipmunk
        Thursday: 4 chipmunks
        Friday: 3 chipmunks
Is this distribution consistent with a Poisson process?

**(10 points)**

c) Hooked on your new exciting hobby, you decide to also observe chipmunks on Saturday, where you observe 9 chipmunks in an hour. What do you conclude?

**(5 points)**

## 2. Reflection

In this course, different teams work on different research projects. Where can you encounter a Poisson process in your project?

**(5 points)**

S06: Homework 6 - Fitting

1. Least Squares Method Afoot

Here, we will analytically work through the Method of Least Squares in a case where it is still possible: fitting a linear function. We make this simple by just fitting a line through the origin such that there's only one fitting parameter (though you can also do this for the more general case where you allow an additional constant as fit parameter too, if you want).

For some distribution $y(x)$ you have $n$ measurements $y_i$ at points $x_i$ which each come with some error $\sigma_i$ ($i=1,2,...,n$). You desire to fit a function $f(x;a_1,a_2,...,a_m)$ to these measurements, i.e. you want to determine the unknown parameters $a_j$ ($j=1,2,...,m$). The *degree of freedom* is defined as $v=n-m$ and should be preferably large. The Method of Least Squares states that the best values $a_j$ are those for which

$$S = \sum_{i=1}^{n} \left(\frac{y_i - f(x_i;a_1,a_2,...,a_m)}{\sigma_i}\right)^2$$

becomes minimal.

a) Interpret this formula.
**(5 points)**

b) In our case, we will fit a straight line with slope $a$. What is the value of $m$? Calculate $S$ in this case.
**(5 points)**

c) To find the parameters $a_j$ for which $S$ is minimal, in general we need to take the partial derivatives $\partial S/\partial a_j$ with respect to the various parameters (partial derivative just means that you e.g., take the derivative with respect to $a_j$ while pretending that all other $a$ are constant). For which value of $a$ does your $S$ become minimal?
**(5 points)**

d) Let's plug in some numbers to make this less abstract. Suppose you measured the following values:

| x | y | $\sigma$ |
|---|---|---|
| 0.1 | 0.3 | 0.1 |
| 0.8 | 1.4 | 0.2 |
| 1.2 | 2.6 | 0.4 |
| 2.6 | 7.1 | 0.9 |
| 3.9 | 8.7 | 0.6 |

What is the slope that you calculate for your fit?
**(10 points)**

2. Reflection

Where can fitting be important in your team's research project?
**(5 points)**

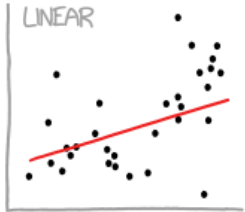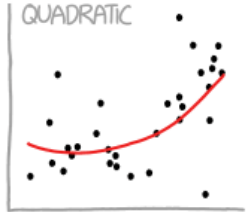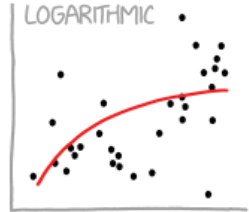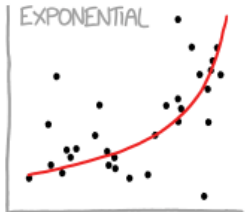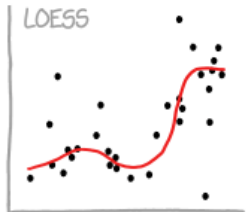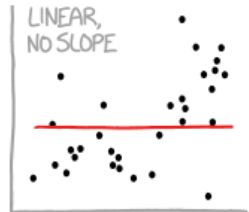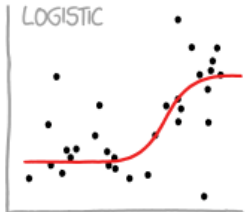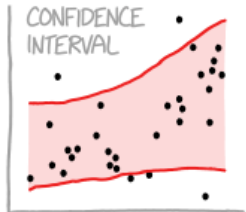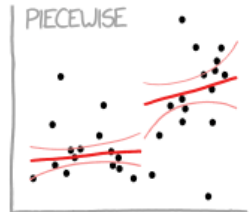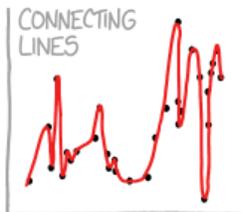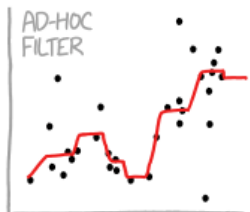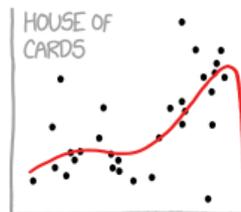

xkcd.com/2048

# S07: Homework 7 - Fermi Problem

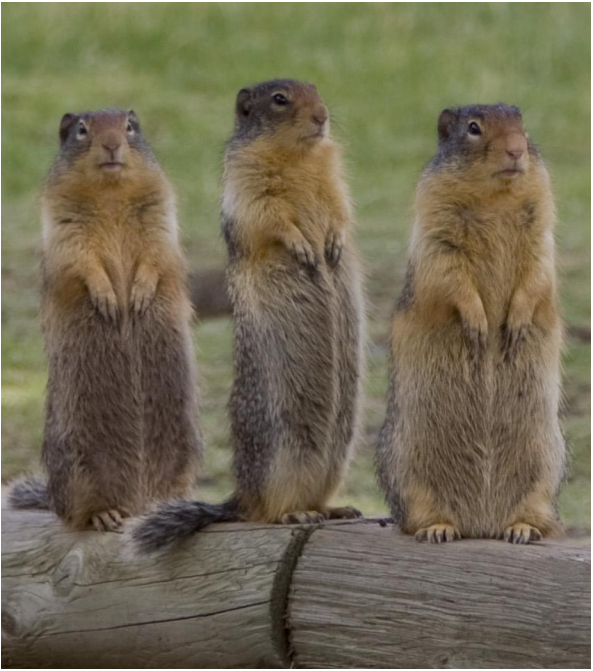

A *Fermi problem* is generally a question that is extremely hard (or even impossible) to answer precisely. For example, "how many grains of sand are there on Earth?". While there is a finite number of grains of sand, the quantity is not precisely knowable. However, we need not give up since we can estimate this quantity through the principle of dimensional analysis. Also, by breaking the problem into many smaller sub-problems, errors in the individual estimates tend to cancel each other out, leading to surprisingly robust results. One well-known example of this method is the Drake-Equation to estimate the number of planets in the universe hosting life. When answering a Fermi question, the method and reasoning that goes into the estimation is more important that the numerical answer. This is an opportunity to use creative and resourceful arguments.

## 1. Fermi Problem

How many chipmunks live on campus?

**(25 points)**

## 2. Reflection

Where can you use this week's lecture content help you with your research project?

**(5 points)**

# S08: Homework 8 - Statistical and Systematic Errors

## 1. Error

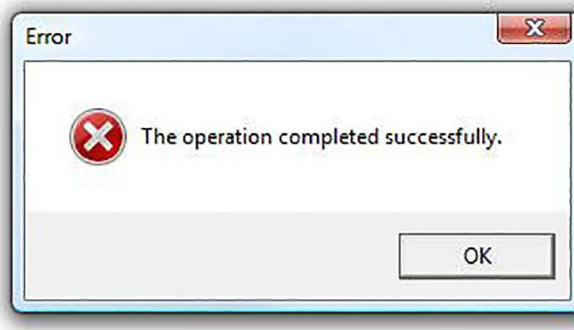

By now you have made various measurements as part of your research. This could can be various rates of events, energy spectra, some performance metric of your analyses, etc. Pick one suitable measurement from your research, such as a single number of events passing all cuts, or a histogram (and thus a collection of numbers in each bin), or some other measurement – really up to you.

a) State the measurement you chose for this assignment: What is it, and how is it measured.
**(5 points)**

b) Calculate a statistical error of your measurement. State how you calculated the error.
**(10 points)**

c) Estimate one systematic error of your measurement. State how you estimated these errors.
**(10 points)**

## 2. Reflection

Here is a homework part where your submission should be different from that of your team members: How does this week's lecture impact you, personally?
**(5 points)**

# S09: Homework 9 - Telescopes

## 1. From ZTF to the LSST

The Zwicky Transient Facility has a limiting apparent magnitude of r < 21 mag. This means that Type Ia supernovae with peak absolute magnitudes of -19.5 mag can be seen as far as 1.2 billion parsecs (1200 Mpc) away, and that the average core collapse supernova peaking at absolute magnitude of - 17 mag can be seen to 400 Mpc. The Legacy Survey of Space and Time to be conducted by the Rubin Observatory will have a limiting apparent magnitude of r < 25 mag, which extends the potential sensitivity distance of Type Ia and core collapse supernovae to 7900 Mpc and 2500 Mpc, respectively.

a) Calculate and compare the potential rate of detections for Type Ia supernovae and core collapse supernovae between the ZTF and LSST surveys. Assume for the rates of supernova explosions in the universe $0.55 \times 10^{-4} yr^{-1} Mpc^{-3}$ for Type Ia supernovae and $1.2 \times 10^{-4} yr^{-1} Mpc^{-3}$ for core collapse supernovae.

**(15 points)**

b) Why does your estimate overestimate the actual observed rate?

**(10 points)**

## 2. Reflection

What is your personal take-aways from this week's lecture and how do they relate to your student path here at the university?

**(5 points)**

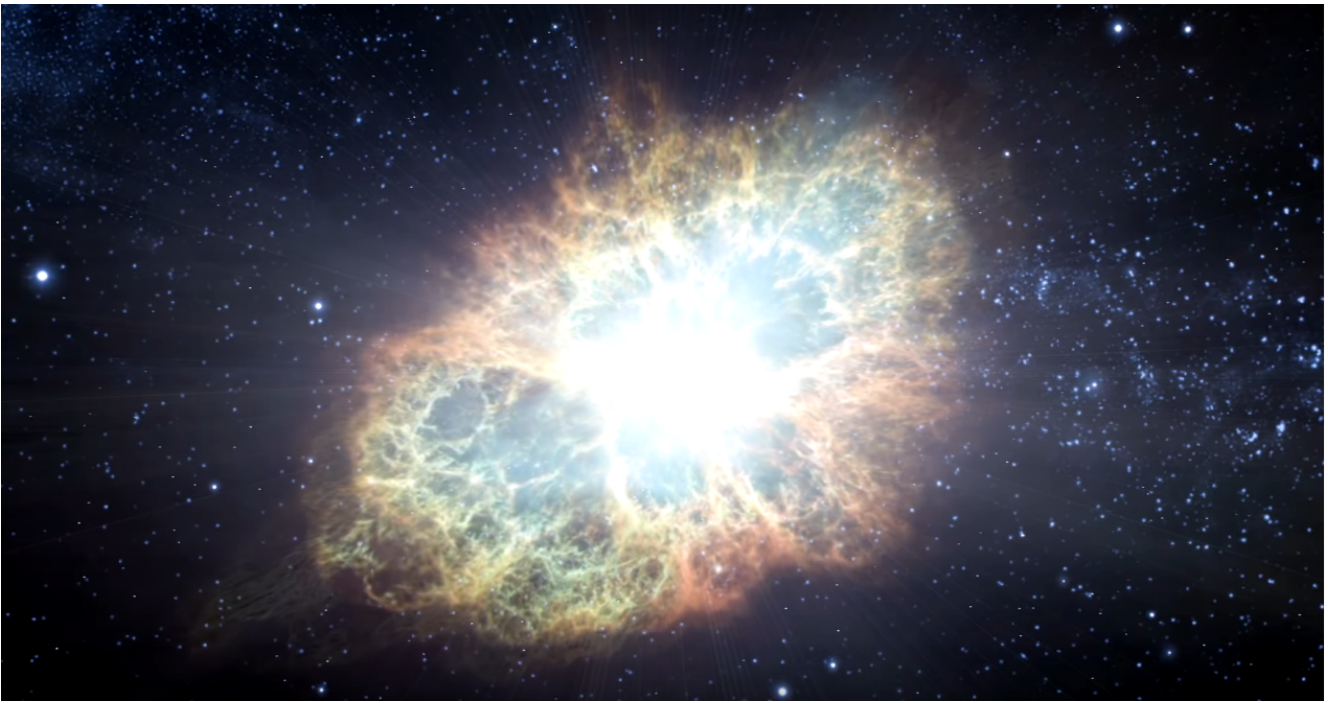

**S10: Homework 10 - Artificial Intelligence**

1. **Artificial Intelligence and You**

Comment on the article on the next pages, reprinted from The New York Times. We are looking for at least three insights that put the article in context with the lecture, and/or relate it to you personally.

**(25 points)**

**2. Reflection**

What is your personal take home message from this lecture?

**(5 points)**

## No Credit Score? No Problem! Just Hand Over More Data.

To determine your risk, start-ups are applying technology to data points as various as your college and the mileage on the used car you want to buy.

By Tara Siegel Bernard

Nov. 29, 2021

For decades, the arbiters of creditworthiness have been two powerful groups: the Big Three credit bureaus, which keep files on roughly 200 million consumers, and score creators like FICO, which turn that raw data into a three-digit key to credit cards, car loans, mortgages and more.

But with tens of millions of consumers left out of traditional credit scoring and the pandemic exposing potential problems in the current system, established players and slick start-ups alike are collecting and crunching all manner of other data to determine who ought to get a loan and how much they should pay.

This so-called alternative credit scoring could have profound effects for consumers, many of them minorities or low-income individuals, who can be asked to hand over more intimate personal information — like their spending habits and details of their college degree — in hopes of getting a loan.

"The box for who gets a conventional credit score is pretty small, and that box hasn't been updated in a while," said Silvio Tavares, the chief executive officer of VantageScore, an established credit scorer that is owned by the big bureaus and is working on adding alternative data to its models. "Data is really a big equalizer."

The efforts to better understand potential borrowers increasingly take two forms, which sometimes overlap. The first involves obtaining cash flow and transaction data from users' bank accounts, a practice that lenders including Kabbage have used. The second involves applying artificial intelligence to broad swaths of information — which may include items already in your credit report, new details such as the mileage on the used car you're buying or perhaps behavior gleaned from your debit accounts — to assess applicants' ability to pay.

Regulators have recently begun discussing both issues, considering and collecting input from the financial industry and others. Officials at the Consumer Financial Protection Bureau have warned that artificial intelligence could amplify risks, including by perpetuating biases against certain borrowers, charging some of them too much or simply making inaccurate predictions. The bureau's director, Rohit Chopra, recently said the new algorithms became "black boxes behind brick walls" when left unchecked.

A deeper understanding of potential borrowers' finances is valuable intelligence for lenders. The roughly 45 million people who have a thin or nonexistent credit history — more than 15 percent of the country's adult population — are a lucrative untapped market.

"FICO is more than 30 years old," said Dave Girouard, chief executive of Upstart, which uses nonfinancial data including the type of job you hold and your level of education to help make credit decisions on personal and auto loans. "It leaves millions of people out in the cold and millions more who pay more for credit than they should."

Upstart's platform is growing rapidly. It has more than 30 lending partners, including Cross River Bank, which made more than 360,000 new loans totaling $3.13 billion in the third quarter, up 244 percent from a year earlier. At least four of those lenders dropped their minimum FICO score requirement altogether.

The company is also something of a regulatory guinea pig: Upstart was the first business to receive a no-action letter from the Consumer Financial Protection Bureau. The letter essentially said the bureau had no plans to take any regulatory action against the company in return for detailed information about its loans and operations.

Though the bureau didn't recreate Upstart's results on its own, it said the company had approved 27 percent more applicants than the traditional model, while the average interest rates they paid were 16 percent lower. For example, "near prime" customers with FICO scores from 620 to 660 were approved about twice as frequently, according to company data. Younger and lower-income applicants also fared better.

Upstart, which also agreed to be monitored by two advocacy groups and an independent auditor, takes into account more than 1,000 data points inside and outside a consumer's credit report. It has tweaked its modeling at times — it no longer uses the average incoming SAT and ACT scores of a borrower's college — but includes the person's college, area of study and employment history. (Nurses rank well, for example, because they're rarely unemployed, Mr. Girouard said.) The amount that borrowers are asking for may also be a factor: If they are seeking more than Upstart's algorithms believe is appropriate, that may work against them.

Other companies work in a similar way, although the methods and data they use vary.

TomoCredit, for example, will issue a Mastercard credit card to applicants — even those with no credit score — after receiving permission to peer at their financial accounts; it analyzes more than 50,000 data points, such as monthly income and spending patterns, savings accounts and stock portfolios. Within two minutes, consumers are approved for anywhere from $100 to $10,000 in credit, to be paid off weekly. On-time payments help build users' traditional credit files and scores.

Zest AI, a Los Angeles company that already works with banks, auto lenders and credit unions, is also working with Freddie Mac, which recently began using the company's tools to evaluate people who may not fit squarely inside traditional scoring models.

Jay Budzik, Zest AI's chief technology officer, said the company went deep into applicants' credit reports, and might incorporate information from a loan application, such as the mileage or potential resale value of a used car. It can also look at consumers' checking accounts.

"How frequently are they getting close to zero?" Mr. Budzik said. "Those things are helpful in creating an additional data point on a consumer that is not in the credit report."

The same methods can also be applied to those who already have a robust credit history, filling out their profiles in real time. Such information became more valuable during the pandemic because credit scores alone may not have picked up signs of stress when borrowers could pause payments on student loans and mortgages.

It can take months for some information to filter into credit scores, said Kelly Thompson Cochran, deputy director of FinRegLab, a nonprofit that tests new technologies in the financial industry. "This can make it particularly difficult for lenders to predict default risk accurately both for applicants who have recently experienced financial difficulties and for applicants who are rebounding from past income or expense shocks," she said.

Established credit scoring and reporting companies are increasingly offering consumers ways to add additional information. The credit bureau Experian's Boost feature allows consumers to pipe in payments on bills from a services like Netflix, Disney+ and their mobile phone provider. The average customer's FICO 8 score — the formula currently used by most lenders — rises 13 points, executives said.

And FICO is piloting a new score, UltraFICO, which augments its traditional model by taking into account — with users' permission — their cash on hand, history of positive balances, and recentness and frequency of banking transactions. FICO estimated the new score can reach 15 million more people.

More information, such as income data or whether you have a 401(k) plan, could be included in future iterations, said FICO's chief executive, Will Lansing. "I think the future of the industry is the consumer taking more control of their data," he said, "and deciding when it will be used and what it will be used for and for what purpose."

Consumer advocates say that's a crucial issue.

While the growing use of transaction data could be a boon to many borrowers, checking and debit accounts contain all sorts of revelatory information, and access to it must remain voluntary, advocates said. Lenders may be looking largely at the broad strokes of your cash flow now, but will they eventually glimpse at where you shop and what types of doctors you visit?

"Credit invisibility is a problem, but some of the solutions or cures can be worse than the disease," said Chi Chi Wu, a staff attorney at the National Consumer Law Center. "It's a high-wire act to make sure this helps more than it hurts."

# S11: Homework 11 - Colliders

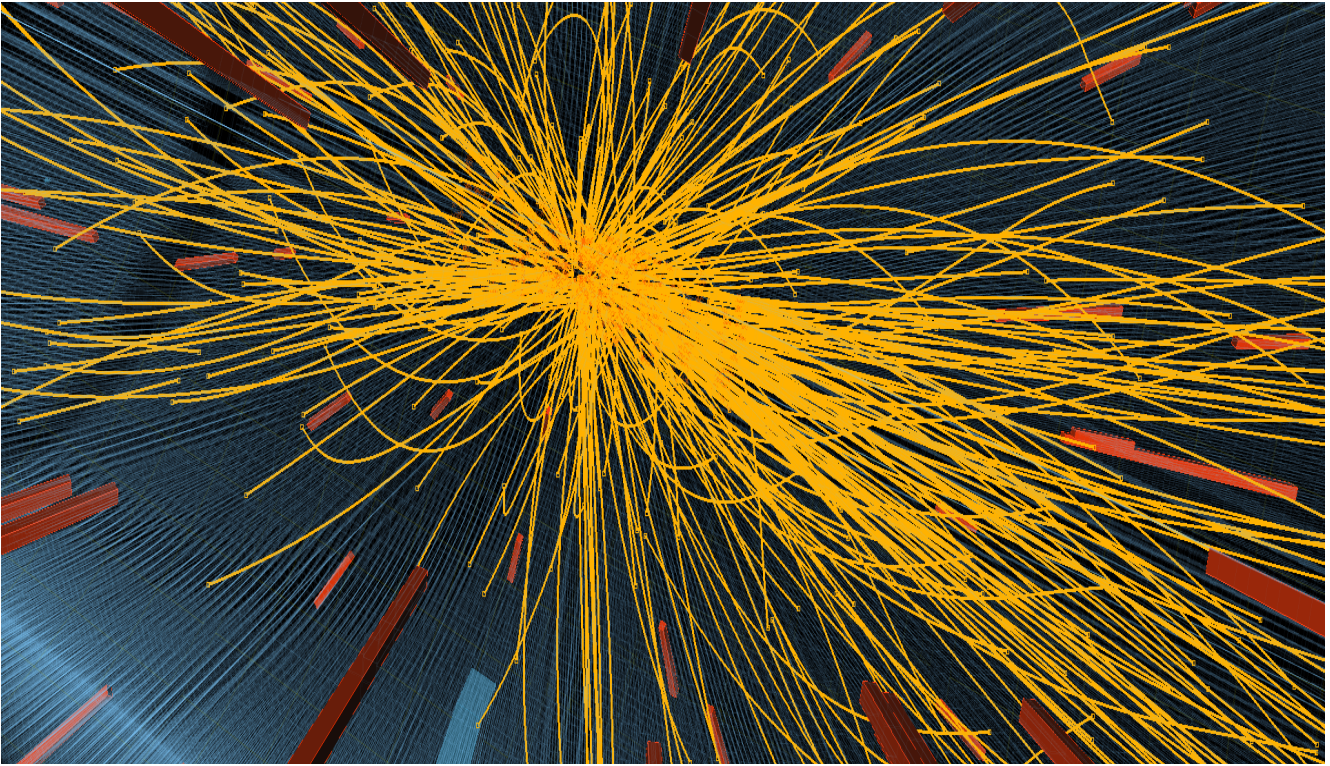

## 1. Collider Experiments
Comment on how (or if at all) you can measure and identify the following particles in a collider experiment such as CMS or ATLAS:
a) Protons
b) Neutrinos
c) b-quarks
d) electrons
e) muons
f) photons
g) Pions
h) Higgs boson
i) top quarks
j) hypothetical dark matter particles

**(25 points)**

## 2. Reflection

What is your personal take home message from this lecture?

**(5 points)**

**S12: Homework 12 - Dark Matter**

**1. Dark Matter Evidence**

In the lecture you have encountered a number of evidence for dark matter, from cosmological to astronomical measurements. Pick one.

a) State the evidence for dark matter you picked: What is the observation, and why does it require dark matter?

**(10 points)**

b) Could this observation be explained by modifying the laws of gravity, instead of introducing additional dark matter?

**(5 points)**

c) From the arxiv, find a recent publication (i.e., from the last couple years or so) that discusses this piece of evidence. What is the novel insight from that publication that wasn't known before?

**(10 points)**

**2. Reflection**

How will this week's lecture content help you with your holiday dinner?

**(5 points)**

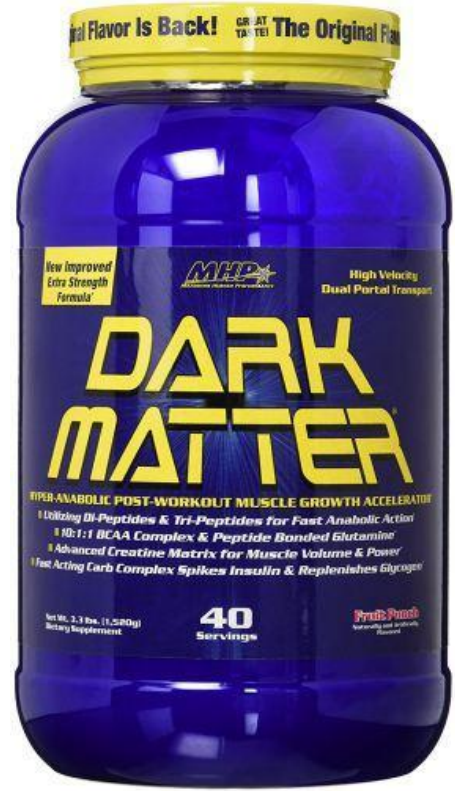

# S13: Physics Presentation Rubric

**Due during the seminar on the day assigned to you.**

Formally present a physics topic. Topics and schedules will be given out by your respective faculty, but please sign up for your topic during the first week of classes.

Limit your presentation to 8 minutes sharp. This will leave another 3 minutes or so for discussions in class. This is typical of a presentation at a meeting of the American Physical Society. Never ever go over your assigned time: your audience won't listen to anything you say after that time.

Note that this assignment carries significant point value, so we recommend you study your topic well, prepare your talk carefully, and rehearse it multiple times. Have a look at the rubric: some aspects are more important to us than others. In particular, note the negative points for failing to meet the 8 minutes.

To research your topic, consider a variety of sources:
- Wikipedia is a great starting point. However, Wikipedia is an encyclopedia and as such does not explain anything. It thus won't be sufficient.
- Talks given by experts e.g., at research conferences, workshops, or summer schools can be a great resource! Be careful to not plagiarize but draw your inspiration from those.
- Amazingly, textbooks do exist. The library is full of them. They explain topics well and put them in context. Use them.
- And/Or find a review article on arxiv.org.
- Add a recent research article to ensure your talk reflects the current state of the art.

Scientific blogs can be a great resource too. Sometimes the pages of papers on the arxiv have links to blogs discussing the paper. They often are more accessible than an original research paper, while giving you the necessary context. Just be aware that they may be simplistic.

For your presentation style and slides, also consider the various points discussed in the lecture, and take into account feedback given to others, in improving your presentation.

**(90 points)**
**(see rubric on next page)**

| Our grading criteria | Not Yet Competent | Competent | Proficient | Blows Me Away |
|---|---|---|---|---|
| Talking Points | Talking points lack relevance to topic or focus<br>**no points** | Talking points could have better related to the topic<br>**6 points** | Talking points are tightly focused and relevant<br>**12 points** | I got a new perspective on things<br>**bonus points** |
| Content | Presentation contains multiple factual errors<br>**no points** | Major facts accurate and generally complete<br>**6 points** | Presentation is accurate with no factual errors<br>**12 points** | I learned something new<br>**bonus points** |
| Completeness | Presentation does not provide adequate depth; key details are omitted or undeveloped<br>**no points** | Missed to present a relevant factoid, major ideas adequately developed<br>**6 points** | Presentation provides good depth and detail; ideas are well developed<br>**12 points** | n/a |
| Current Content | Could have been help a century ago<br>**no points** | Barely mentions a current detail<br>**6 points** | At least a slide dedicated to current research<br>**12 points** | Multiple slides with recent news on the topics<br>**bonus points** |
| Clarity | Several parts of presentation are wordy or unclear<br>**no points** | Some parts of presentation are wordy or unclear<br>**6 points** | Presentation is clear and concise<br>**12 points** | n/a |
| Organization | Presentation is jumbled up, transitions are lacking<br>**no points** | Story flows but somehow is told backwards<br>**3 points** | Presentation order flows logically with smooth transitions<br>**6 points** | n/a |
| Documentation | Visual aids are missing or inadequate<br>**no points** | Adequate message support provided for key concepts by facts and visual aids<br>**3 points** | Effective message support provided in the form of facts and visual aids<br>**6 points** | n/a |
| Slide Design | Poor contrast, too small fonts, crowded slides, no page numbers<br>**no points** | Slides don't distract but support the speaker's message<br>**2 points** | Good visuals, large fonts, little text, page numbers<br>**4 points** | n/a |
| References | No references provided<br>**no points** | Sourcing is generally adequate<br>**1 point** | References appropriate and current<br>**2 points** | n/a |
| Delivery | Low energy; pace too slow or fast; poor diction<br>**no points** | Adequate energy, pace and diction<br>**1 point** | Good volume and energy; proper pace and diction<br>**2 points** | TV-ready delivery<br>**bonus points** |
| Grammar | Presentation contains major grammar/usage errors; sentences are long, excessive jargon<br>**no points** | Presentation has no serious grammar errors; sentences are mostly jargon-free<br>**1 point** | Sentences are free of jargon and easy to understand<br>**2 points** | n/a |
| Interactions | No eye contact with audience<br>**no points** | Fairly good eye contact with audience<br>**1 point** | Good eye contact with audience<br>**2 points** | Getting the spark across<br>**bonus points** |
| Appearance | Distracting gestures or posture; unprofessional appearance<br>**no points** | Few or no distracting gestures; solid appearance<br>**1 point** | No distracting gestures; professional appearance<br>**2 points** | n/a |
| Duration | Presentation over 8.5 or under 7 minutes<br>**-10 points** | | Finished within 450-490 seconds<br>**2 points** | n/a |
| Questions | Poor listening skills: inability to answer audience questions<br>**no points** | Adequate answers to audience questions<br>**1 point** | Answers to audience questions with authority and accuracy<br>**2 points** | Did you write a PhD on this topic?<br>**bonus points** |

**S14: Research Briefing Rubric**

**Due during the seminar on the day assigned to you.**

Research is often communicated in rather informal settings to small groups, be it your experiments' collaboration in a telecon, or your faculty's research group in a weekly team meeting. This research briefing assignment simulates such environments to practice the communication of progress and plans for your team research project. As we ask your team to brief your colleagues twice this semester, this assignment exists twice in Blackboard.

Remember to introduce your fellow students to your research project before going into details. Keep the level appropriate to their prior knowledge – just put yourself in their shoes. Present your progress since the last research briefing. Also give ideas on how you intend to go forward.

Consider the time assigned to your meeting. You should allow for frequent interruptions, questions, and discussions. It is thus perfectly adequate to directly talk everybody through a Jupyter note. Just make sure that the note is well designed, clearly described, flows well, and has high quality plots. Then a formal presentation isn't needed (and in fact could be counter-productive to a detailed discussion). It may happen that you don't get through your entire note and that is fine too, as this is not a final presentation of a concluded project, but rather a status report on work in progress.

Part of this assignment will be matrix graded, so each team member should write a simple email to your teaching assistant about the relative contributions of your fellow team members (excluding you), which should add up to 100%.

**(90 points)**
**(see rubric on next page)**

| Our grading criteria | Not Yet Competent | Competent | Proficient | Blows Me Away |
|---|---|---|---|---|
| **Introduction to Research** | No introduction<br><br>**no points** | Some introductions but insufficient for some that haven't worked on that exact topic.<br><br>**10 points** | Clearly gets everybody from their prior knowledge up to speed with what this briefing is going to be all about<br><br>**20 points** | n/a |
| **Research Presentation** | Unclear oral presentation, insufficient and/or very low-quality visual material<br><br>**no points** | Basic information is present but note misses e.g., explanations, relevant details, axis labels, etc.<br><br>**15 points** | Clear flow of presentation and visuals/note which includes high quality figures<br><br>**30 points** | This should have been taped to be preserved for eternity…<br><br>**bonus points** |
| **Research Results** | No progress made since last presentation<br><br>**no points** | Some progress has been made even though no new insights yet<br><br>**10 points** | Significant progress with new insights into this research project<br><br>**20 points** | Nobel-prize worthy discovery<br><br>**bonus points** |
| **Plans going forward** | No plan<br><br>**no points** | Only vague ideas or ideas not helpful to the research project<br><br>**5 points** | Clear ideas on how to drive the research to a conclusion<br><br>**10 points** | Some really clever ideas<br><br>**bonus points** |
| **Discussion** | No contribution to discussion<br><br>**no points** | Some lack of confidence or contributions to the discussion<br><br>**5 points** | Discussion and questions handled with confidence and insight<br><br>**10 points** | In charge of entire group and discussion<br><br>**bonus points** |
| | | | Total 90 points. 1/3 of those distributed based on matrix-grading. | |

**S15: Team Preparedness Rubric**

For any team-based research it is crucial that you recognize the unique strengths of each team member and exploit them whenever possible. Further, the review of any research proposal will look at the qualifications of the proposing team to conduct the proposed research. For example, the National Science Foundation explicitly asks their referees: "How well qualified is the individual, team, or organization to conduct the proposed activities?". Write a text addressing this review criterion. You are welcome to recycle all or parts of this assignment for the assignment later this semester where you will write a full proposal.

Argue what each team member brings to the table to conduct your semester-long research project. For the sake of this assignment, this can include prior knowledge, your majors, interests, personality traits, etc. Describe how the team members will contribute to this research project. Synergy is a huge buzzword whenever somebody is giving out funding, so yes, synergies should be apparent and spelled out clearly.

**Use Latex**

Latex is widely used in physics, math, and related fields, where almost all scientific articles and theses are written using this system. Believe me when I say that you do not want to write a 100-page thesis with dozens of figures, cross-references, and references, using Word or similar WYSIWYG editors. The basic idea of Latex is to separate content from presentation. With Latex, you simply use a professional template for the design and layout, and then never worry about things like font sizes for titles, or page numbers etc., ever again. This saves a huge amount of headache in the long run and is extremely robust. You could use any plain-text editor to write a document, and then compile it into a pdf. But even easier, simply use it through the convenient online editor www.overleaf.com which also permits collaborative editing (as in e.g., google docs). We strongly advise against starting a document from scratch though and will provide a template for you to start with; you should have an easy time with that. If not, remember to never hesitate to ask Rafael or your TA for help.

**Matrix Grading**

Submit only one submission for your entire team to Brightspace. This assignment will be matrix graded, so in addition, each team member should write a simple email to your teaching assistant about the relative contributions of your fellow team members to this homework (excluding you), which should add up to 100%.

**(30 points)**
**(see rubric on next page)**

| Our grading criteria | Not Yet Competent | Competent | Proficient | Blows Me Away |
|---|---|---|---|---|
| **Presentation of team member** | The qualifications and contributions of this team member are not discussed. | This team member is discussed, but missing unique qualifications or connections to the research project. | At least one unique qualification and its relevance to the research project is clearly developed. | n/a |
| **Member 1** | no points | 2 points | 4 points | |
| **Member 2** | no points | 2 points | 4 points | |
| **Member 3** | no points | 2 points | 4 points | |
| **Member 4** | no points | 2 points | 4 points | |
| **Team Preparedness** | The reader is left doubtful that the team is capable to perform the proposed research. <br><br> no points | Partial arguments are made that some team members are prepared for some of the proposed research. <br><br> 3 points | A clear argument is made how each team member will contribute to the proposed research. <br><br> 6 points | A convincing case is made that this is the dream team for the proposed research. <br><br> bonus points |
| **Synergies** | No synergies are presented. <br><br> no points | Some synergy may be present, but the reader is left unconvinced of the relevance to the research project. <br><br> 3 points | Synergies are clearly present and can be expected to amplify this team's capabilities. <br><br> 6 points | Synergies exponentially amplify this team's capabilities. <br><br> bonus points |
| **Language** | Many grammatical, spelling, or punctuation errors distract the reader from the content. <br><br> no points | A few grammatical, spelling, or punctuation errors interrupted the flow of reading by catching the reader's attention. <br><br> 1 point | The text is easy to read and contains no grammatical, spelling or punctuation errors. <br><br> 2 points | n/a |

# S16: Literature Review Rubric

Any research project starts with a survey of what is known. Thus, write a literature review for your team's research project. You will want to have a minimal broad introduction to your experiment and topic. Then get more detailed as you drill down to the particular research question of your team.

You are writing a scientific literature review as one would find in a proposal or a publication; thus, information that can readily be found on wikipedia or introductory textbooks need not be referenced at all. The best references are original, peer-reviewed research publications. Find them e.g., on www.arxiv.org.

You are welcome to recycle all or parts of this assignment for the later team assignment of the complete research proposal. There are no minimum nor maximum limitations on your document but remember to write in a reviewer-friendly way. Again, we strongly recommend using Latex (and in particular Overleaf) for this assignment.

**Bibliography in Latex**

Citations are particularly easy using Latex. Say you want to cite the 2007 paper *The Importance of Being First: Position Dependent Citation Rates on arXiv:astro-ph* by Dietrich. While there are many ways to do so, by far the easiest one in the long run is to use the Bibtex plugin with Latex. For example, this specific paper is here https://arxiv.org/abs/0712.1037 as open access version, as is essentially any physics paper since the late 1990s. There, you can e.g. click on the "INSPIRE HEP" link to find its Digital Object Identifier DOI 10.1086/527522 which gets you to the published version if you care, but that's often behind a pay wall. Now, also on INSPIRE HEP, you can click on "Bibtex" to find the little sniplet text that you can copy directly into your bibliography.bib file, for use with latex. Super convenient, and since most in our fields use this system, you can even find scripts online that automatically fill your bibliography.bib file from just the INSPIRE HEP article tags in your .tex document. That tag would be Dietrich:2007wu in this case, so the suggestion is to simply leave that as is.

Again, your TA will be happy to help you using this professional writing software. Remember to write in a maximally reviewer-friendly way, so citing papers as [Dietrich 2007] is better than citing as a simple number [4].

Submit only one pdf for your entire team to Brightspace. This assignment will be matrix graded, so each team member should write a simple email to your teaching assistant about the relative contributions of your fellow team members (excluding you), which should add up to 100%.

**(30 points, matrix graded)**
**(see rubric on next page)**

| Our grading criteria | Not Yet Competent | Competent | Proficient | Blows Me Away |
|---|---|---|---|---|
| Introduction | No introduction to this research question. **no points** | Some introduction is present but not at the level adequate for a research paper or proposal. **3 points** | A clear introduction guides the reader from the field to the experiment to the particular research. **6 points** | n/a |
| Review | Review is too brief, lacks significant detail, and/or is not related to the particular research question. **no points** | A review that is superfluous in parts or lacks the required detail to properly assess the body of existing research. **5 points** | A thorough review of knowledge related to the specific research question. **10 points** | Publication-ready material. **bonus points** |
| References | Multiple relevant statements missing references, and absence of key publications. **no points** | A small number of references, or a lack of current work. **5 points** | Many relevant and, where appropriate, current peer-reviewed articles. **10 points** | Includes highly relevant citations previously unknown to the instructor. **bonus points** |
| Presentation | Substantial issues with inappropriate formatting or length. **no points** | Distracting features in the formatting **1 point** | Neat presentation with smooth readability **2 points** | n/a |
| Language | Many grammatical, spelling, or punctuation errors distract the reader from the content. **no points** | A few grammatical spelling, or punctuation errors interrupted the flow of reading by catching the reader's attention. **1 point** | The text is easy to read and contains no grammatical, spelling or punctuation errors. **2 points** | n/a |

**S17: Preliminary Results Rubric**

In the previous two parts of your semester-long project to write a research proposal, you have already demonstrated to any reviewer that your team has the background to successfully conduct your proposed research, and you have shown that you are familiar with the state of the art through your literature review. Now it is time to drive that message home by showing that you already did some preliminary work; that could be collected some data, conducting some analysis, or making some estimates to the feasibility of the research you will ultimately propose.

Show what you have already accomplished towards achieving your goal. Thus, demonstrate that what you are going to propose can be expected to lead to a favorable outcome.

This part of your proposal should definitely include some figures or graphs that you made yourself.

Submit only one pdf for your entire team to Brightspace.

We will again matrix grade part of this assignment, so each team member should write a simple email to your teaching assistant about the relative contributions of your fellow team members (excluding you), which should add up to 100%.

**(60 points)**
**(see rubric on next page)**

| Our grading criteria | Not Yet Competent | Competent | Proficient | Blows Me Away |
|---|---|---|---|---|
| **Statement of Problem** | No problem identified within the scope of this class. **no points** | The chosen research question needs some polishing or focus but is a sound basis for this semester project. **2 points** | The chosen research question is clearly articulated and correctly phrased in the context of present knowledge. **5 points** | The research question has the potential to lead to a publication sometime down the road. **bonus points** |
| **Prior Work** | Only rudimentary plots e.g., taken straight from examples, inadequate axis labels, no focus on relevant parameters. **no points** | Ability to make plots which may however still have some issues with axis labels or ranges; demonstrate understanding of the relevant parameter space. **10 points** | Demonstrated proficiency with manipulating data, producing quality plots. **20 points** | The team already made significant strides towards realizing new insights. **bonus points** |
| **Proposed Work** | Simplistic plan going forward. **no points** | While the planned work is outlined, it is vague in the approach taken, the analyses that need to be done, and/or the plots to be made. **10 points** | A clear presentation of the work ahead of the team and how it relates to answering the research question. **20 points** | Creative ideas, new viewpoints, novel connections, … **bonus points** |
| **Organization** | No timeline and no distribution of tasks across team members. **no points** | Some organizational thoughts are presented but lack either a clear schedule or a clear distribution of tasks across team members. **5 points** | A feasible timeline is presented together with a sound distribution of tasks across the team members. **10 points** | Meaningful implementation of organizational charts, gannt charts, etc. **bonus points** |
| **Language** | Many grammatical, spelling, or punctuation errors distract the reader from the content. **no points** | A few grammatical spelling, or punctuation errors interrupted the flow of reading by catching the reader's attention. **3 points** | The text is easy to read and contains no grammatical, spelling or punctuation errors. **5 points** | n/a |
| **Teamwork** | | | Total 60 points, half of which are matrix-graded | |

# S18: Proposed Work Rubric

A full proposal to a funding agency can be very involved and forms the basis to support research programs all over the world. The competition for funding support is fierce (with success rates commonly being below 15%), so much advice exists on how to write good proposals. Any funding proposal needs to address your team preparedness, a thorough review of prior knowledge (i.e., pertinent literature), and preliminary results towards your proposed research goal. Since we already covered those parts, we can keep it simple here and focus on three main aspects:

- The research question. Formulate the problem at hand, what you plan to do and why it is interesting. This is future-oriented, i.e., not a statement of what you already did, but what you would like to investigate going forward.
- A plan for going forward. Sketch out this plan, provide a schedule, and distribute roles across your team. For concreteness, assume that all team members would continue next semester, and that both your team and your research topic would remain the same. (Either of these may or may not be the case but we will ignore that here). Thus outline your proposed work across the next semester.
- Pull your previously written parts and these new sections together into one coherent narrative. Write in a maximally reviewer-friendly way to put your reviewers (aka your TA) in a good mood as you need to convince them to support your research.

Typical proposals to the National Science Foundation are for three-year research projects and can carry budgets ranging from tens of thousands of dollars to a million or more. And yet the default NSF proposal is at most 15 pages long! So, given that our requested dollar amount is much smaller, we will restrict this proposal to at most 10 pages.

Use your previous proposal parts as a starting point and build the proposal from there. Direct copy and paste is perfectly alright, but of course you are free to modify or re-write as you see fit.

We will again matrix grade part of this assignment, so each team member should write a simple email to your teaching assistant about the relative contributions of your fellow team members (excluding you), which should add up to 100%.

**(120 points)**
**(see rubric on next page)**

| Our grading criteria | Not Yet Competent | Competent | Proficient | Blows Me Away |
|---|---|---|---|---|
| **Statement of Problem** | No problem identified within the scope of this class.<br><br>**no points** | The chosen research question needs some polishing or focus but is a sound basis for this semester project.<br><br>**5 points** | The chosen research question is clearly articulated and correctly phrased in the context of present knowledge.<br><br>**10 points** | The research question has the potential to lead to a publication sometime down the road.<br><br>**bonus points** |
| **Prior Work** | Only rudimentary plots e.g., taken straight from examples, inadequate axis labels, no focus on relevant parameters.<br><br>**no points** | Ability to make plots which may however still have some issues with axis labels or ranges; demonstrate understanding of the relevant parameter space.<br><br>**20 points** | Demonstrated proficiency with manipulating data, producing quality plots.<br><br>**40 points** | The team already made significant strides towards realizing new insights.<br><br>**bonus points** |
| **Proposed Work** | Simplistic plan going forward.<br><br>**no points** | While the planned work is outlined, it is vague in the approach taken, the analyses that need to be done, and/or the plots to be made.<br><br>**20 points** | A clear presentation of the work ahead of the team and how it relates to answering the research question.<br><br>**40 points** | Creative ideas, new viewpoints, novel connections, …<br><br>**bonus points** |
| **Organization** | No timeline and no distribution of tasks across team members.<br><br>**no points** | Some organizational thoughts are presented but lack either a clear schedule or a clear distribution of tasks across team members.<br><br>**10 points** | A feasible timeline is presented together with a sound distribution of tasks across the team members.<br><br>**20 points** | Meaningful implementation of organizational charts, gannt charts, etc.<br><br>**bonus points** |
| **Language** | Many grammatical, spelling, or punctuation errors distract the reader from the content.<br><br>**no points** | A few grammatical, spelling, or punctuation errors interrupted the flow of reading by catching the reader's attention.<br><br>**5 points** | The text is easy to read and contains no grammatical, spelling or punctuation errors.<br><br>**10 points** | n/a |
| **Teamwork** | | | Total 120 points, half of which are matrix-graded | |

**S19: Pre-Tutorial**

**One submission per research team, due to your TA at the start of your team's tutorial in whichever form you agree with your TA.**

What did your team do, where did you get stuck, and what are your plans going forward?

**(10 points)**

**S20: Homework - Schedule your Weekly Tutorial**

**Deliverable: Simply communicate this with your TA and team members.**

Schedule time and place of your groups' weekly tutorial with your TA. Also identify a time and place for your regular team meetings. Communicate both with all team members and your TA.

**(10 points)**

## S21: Matrix Grading Example

This assignment is worth **50 points**

In Team Z, members are Ann, Barb, Chris, and Dee:
- Ann feels Barb didn't pull her weight but thinks Chris and Dee contributed about equally:
- Barb appreciated Ann really pushing for this, and thinks Dee did a bit less than Chris:
- Chris thinks Ann really helped by taking the lead on this, with Barb doing a bit less than Dee:
- Dee had the best interaction with Ann and Barb but realizes Chris had a key idea:

| Ann | Barb | Chris | Dee | Sum |
|---|---|---|---|---|
|  | 20% | 40% | 40% | 100% |
| 70% |  | 20% | 10% | 100% |
| 60% | 15% |  | 25% | 100% |
| 30% | 30% | 40% |  | 100% |

The weights assigned by each team member need to add up to 100%.

But at the end, each student can get more or less than 100% of the available points:

| Ann | Barb | Chris | Dee | Sum |
|---|---|---|---|---|
| 160% | 65% | 100% | 75% | 400% |

Now, the TA, found the submission of Team Z to be worth **40 points.**

Thus, the final scores of each team member is calculated to be:

| Ann | Barb | Chris | Dee | Sum |
|---|---|---|---|---|
| 64 | 26 | 40 | 30 | 160 |

Note that Ann got 64 points out of a 50 point assignment, and that Barb should contribute more to the team's success.

Debrief all graded assignments in your team with your TA. Constructive criticism is key. Talk with the TA how a higher score could have been achieved.

**S22: Homework -Feedback**

**1. Feedback Please**

We highly value your feedback. What should we change about this course? What is good and should be kept? What is bad, and how would you do it differently?

Please help us improve this course. We won't give out points but highly welcome your opinion!

**(10 bonus points)**
**(full points independent of what you submit, as long as it comes in before the deadline)**

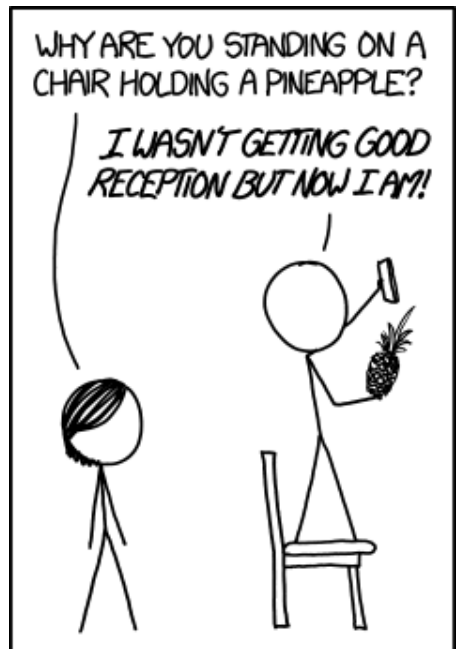